\documentclass[twocolumn,prd]{revtex4-1}
\usepackage{graphicx}
\usepackage{flafter}
\usepackage{calc}

\begin{document}
\title{\textbf{Surface tension with Normal Curvature in Curved Space-Time}}
\author{Himanshu kumar}
\email{hman\_19@hotmail.com}
\author{Sharf Alam}
\author{Suhail Ahmad}
\affiliation{Department of Physics, Jamia Millia Islamia, New Delhi, India}


\begin{abstract}
With an aim to include the contribution of surface tension in the action
of the boundary, we define the tangential pressure in terms of surface
tension and Normal curvature in a more naturally geometric way. For a thin
shell approximation of a static spherically symmetric surface and for
weak and slowly varying fields, the negative tangential pressure ταβ
is chosen to be analogous to Sαβ , where Sαβ is the classical surface
tension. First, by a suitable choice of the enveloping surfaces, we show
that the negative tangential pressure is independent of the four-velocity
of a very thin hyper-surface. Second, using suitable definition of the
normal curvature for such a surface layer, we relate the 3-pressure of
a surface layer to the normal curvature and the surface tension. Third,
using the fact that the tangential pressure on the surface layer is
independent of the four-velocity and a central force interaction, we
relate the surface tension Sαβ to the energy of the surface layer. Four,
we show that the delta like energy flows across the hypersurface will be
zero for such a representation of intrinsic 3-pressure. Five, for the weak
field approximation and for static spherically symmet- ric configuration,
we deduce the classical Kelvin’s relation between surface tension,
pressure difference and mean curvature from this sort of representation
of negative tangential pressure ταβ in terms of surface tension Sαβ
and the normal curvature. Six, using the representation of tangential
pressure in terms of surface tension and normal curvature, we write a
modified action for the boundary having contributions both from
surface tension and normal curvature of the surface layer. Also we propose
a method to find the physical action assuming a reference background,
where the background is + not flat. (The gμν or just gμν has been
chosen to represent the metric coefficent of the − hypersurface of V+
space which is time-like surface layer here. The gμν represents the
metric coefficient of the space like hypersurface of V− space.)
\end{abstract}


\maketitle

\section{Introduction}
\label{intro}
In General relativity,an energy - momentum tensor concentrated on a time - like hypersurface is called a surface layer.Einstein's equations relate it to the jumps of the second fundamental form equivalent to the extrinsic curvature.The algorithms to calculate the same has been given in\cite{lanczos1924}-\cite{Kuhar1968}.
A Surface layer has no component in the normal direction ,otherwise the delta like character would be destroyed.Hence an ideal fluid with non - vanishing pressure cannot be concentrated on an arbitrarily thin region therefore in most cases this layer is composed of dust.
Additionally, an intrinsic 3- pressure has been taken into account in\cite{Einstein1939}\cite{Gott1980}. 
The absence of radial pressure  has been illustrated by assumption of identical particles moving on circular orbits \cite{Schmidt1983},\cite{Lake1979} and \cite{Maeda1983}.
The intrinsic 3-pressure and surface tension are of both dimensions force per unit length.
The 3-pressure(equal to tangential pressure) of a surface layer has been related to surface tension in \cite{Schmidt1984}.
In \cite{Schmidt1984} ,negative tangential pressure depends on the four-velocity of the hypersurface.We make the tangential pressure independent of four-velocity on the surface layer.
This we do by a suitable choice of two enveloping surfaces $V_{+}$ and $V_{-}$ embedding the surface layer.
$V_{+}$ has time-like hypersurface and $V_{-}$ has space-like hypersurface.
The neighborhood points of the hypersurface of $V_{+}$ represent the outermost regions of the surface.$V_{+}$  is a hyperboloid sheet curved upwards.(see sec.~\ref{sec:4}).
The hypersurface of $V_{+}$ has positive normal curvature.The $V_{-}$ space is a hyperboloid sheet curved downwards.
The points  in the neighbourhood of $V_{-}$ are inside the surface.The hypersurface of $V_{-}$ space is the inner surface of the patch on the surface layer.
This hypersurface only overlaps a thin region of the hypersurface of $V_{+}$.These two hypersurfaces don't intersect,so there are no inner boundaries \cite{Hawking1995}.
The hypersurface of $V_{-}$ has negative normal curvature.A discussion of surface layer based on the second fundamental form has been done in \cite{Voohees1972}.
The  second fundamental form must not be equal in the two enveloping manifolds having a common boundary.
It must have opposite signs for the surface layer to exist.
 The surface layer is the time-like hypersurface of $V_{+}$ space.This description of the surface  makes the tangential pressure on the surface layer independent of the four-velocity.
 The points on the surface layer  become stationary.
 We therefore consider the surface tension of the surface  layer as a central force interaction.For this choice of a surface, surface tension  arises only from the internal interactions of the points on surface layer.
 Therefore we assume the work done by the stretching force of surface tension  to be stored as energy of the surface layer.
 There can be different phenomenological representations of $\tau_{\alpha\beta}$,the negative tangential pressure to arrive at the classical Kelvin' relation. Here we have used the definition of normal curvature H  of the time-like hypersurface of $V_{+}$ space as the analogue of Mean curvature,represented by H in kelvin's formula \cite{Kelvin1858}.
 Normal curvature has been preferred since it only depends on the point on the patch of the surface layer where it is calculated.Also it only depends on the sense of the normal of  the enveloping spaces of the hypersurface.The sum of the normal curvatures of surface is defined as the product of the \textbf{m.etric} and the \textbf{second fundamental tensor}.Note that the second fundamental tensor is equivalent to the extrinsic  curvature $K_{\mu\nu}$ of the hypersurface.We know that the extrinsic curvature  defines how a hypersuface bends in the space-time in which it is embedded \cite{Wald1984}.The dimension of the enveloping space is only one greater than that of the hypersurface.This ensures that there will only be one unique linearly independent normal curvature of the hypersurface of $V_{+}$ space which is the surface layer here. We calculate the pressure difference across the surface layer.For this we use the change in the divergence of the unit normal in the two enveloping spaces i.e.$V_{+}$ and $V_{-}$. \\
 The paper proceeds as follows:Section 2 contains Kelvin's relation for non - relativistic surface tensions discussed in \cite{Schmidt1984}.Also here we give a brief of the representation of tangential pressure in \cite{Schmidt1984}.
 We justify  our representation of the same.In section 3,  we discusses the definition of normal curvature  of hypersurface.Also we define the tangential pressure  in terms of the normal curvature and Surface tension tensor. 
 In Section 4,we establish the absence of delta -like energy flows across the surface  ,for our choice  of phenomenological representaion of negative tangential pressure on the surface layer which is $\tau_{\alpha,\beta}$ here. 
 Also we derive a relationship between the surface tension tensor  of the hypersurface and the energy- momentum tensor of the enveloping surface.Here we also prove that for weak field approximation  and atleast for static spherically symmetric configuration of constant radii ,our expression for pressure difference across the surface layer reduces to the Classical Kelvin's relation between pressure difference across the surface,surface tension and the mean curvature.In the last part of this section,we write a modified action for the boundary.This action has contribitions both from the normal curvature as well as the surface tension term of the hypersurface,which represents the surface layer.We also propose a method to calculate the physical action assuming a reference background,where the background is not flat. Appendix I we have illustrated the use of Gaussian Normal cordinates to represent the enveloping surfaces of the hypersurface.Also we show a diagrammatic representation of our manifold.In Appendix II we  show that the negative of the divergence of the unit normal in the $V_{n + 1}$ enveloping  space is  the mean curvature of the hypersurface. Therefore mean curvature is negative of the expression for normal curvature \cite{Eisenhart1927}.
 \section{Classical Surface Tensions }
 \label{sec:1}
 For a drop of liquid in vacuum,whose equilibrium configuration is a spherical one,Kelvin's relation \cite{Kelvin1858} between surface tension S,pressure difference(outer minus inner pressure)$\Delta P$ and mean curvature H(where H $=\,1/R$ for a sphere of radius R) 
 reads:
 
 \begin{equation}\Delta P=-2HS
\end{equation}
Here S is a material dependent constant and an energy $S.\Delta A$  is needed to increase the surface area by $\Delta A$.This supports the description of $-S$ as a kind of intrinsic pressure \cite{Schmidt1984}.In addition for a non - spherically symmetric surface,the mean curvature H in eq(1) may be obtained from the principal curvature radii $R_{1}$ , $R_{2}$ by means of the relation \begin{equation}
H=\frac{1}{2R_{1}} + \frac{1}{2R_{2}}\end{equation}In \cite{Schmidt1984},they have defined the negative tangential pressure on surface layer as:
\begin{equation} \tau_{\alpha \beta}=-S(g_{\alpha \beta} + u_{\alpha}u_{\beta})\end{equation} where $u^{\alpha}u_{\alpha}=-1$.Here and $g_{\alpha \beta}$ is the metric coefficent of the hypersurface,$u^{\alpha}$ is the four-velocity parallel to the hypersurface and S is the classical surface tension.They have defined $\Delta P$ which is the pressure difference across the hypersurface as :\begin{equation} \Delta P =-2HA=\frac{1}{2}(k_{\alpha\beta}^{+} + k_{\alpha\beta}^{-})\tau^{\alpha \beta}\end{equation}\\where $k_{\alpha\beta}^{+}$ and $k_{\alpha \beta}^{-}$ are the second fundamental tensor  of the enveloping spaces on both sides of the hypersurface.$H$ is the mean curvature of a boundary defined on the hypersurface.They define the boundary at a fixed moment by using  a curl free four-velocity.That is their four-velocity is the gradient of some scalar.They  then define the principal curvature directions inside the above defined boundary.The effect of pressure anisotropy on surface tension of compact stars has been studied in \cite{Maharaj2007}.To simplify things , we have used the normal curvature of the surface layer.We use normal curvature to define tangential pressure in terms of the surface tension tensor to give it a more natural geometric meaning.A scalar-invariant form of curvature has been used to simulate free surface flows with surface tension  in \cite{Alexander2006}.The preference of scalar-invariant form of curvature over mean curvature has also been suggested in \cite{David2011}.Also see the representation of surface tension force and curvature in \cite{Zemach1992},\cite{Yakovenko2011},\cite{Kothe2006},\cite{Zaleski1999},\cite{Yuriko2002}. Note that the surface layer here is the hypersurface of the $V_{+}$ space. Recall that the neighbourhood points of $V_{+}$ enveloping space represent the outermost regions of the source (see fig.\ref{hyperbole}). We use the fact that the normal curvature as per its definition only depends on the point of the hypersurface at which it is defined and the sense of the normal.Also if we assume two enveloping spaces on opposite sides of the surface layer. The  negative of the divergence of the unit normal in these enveloping spaces  represents the mean curvature of the surface layer.We don't need any constraint equation on the hypersurface as in \cite{Schmidt1984}.By suitable description of the surface, we make the surface layer independent of four-velocity.The tangential pressure on the hypersurfaces of $V_{+}$ and $V_{-}$ is negative.This is because a very thin hypersurface is assumed to be massless \cite{Onokondo1960}.The normal curvature  is defined  at a point P on the surface layer.Recall that the surface layer here is the time-like hypersurface of $V_{+}$ space.It is on a patch of surface composed of overlap of hyperboloid  of two sheets.The sheet which is curved upwards at the point P is the $V_{+}$ space. The points in the neighborhood of which lie on the outermost regions of the source.The other sheet curved downwards at P  is the $V_{-}$ space. The neighbourhood points of which lie inside the source. $V_{-}$ space has a space-like hypersurface.We thus assume that the negative tangential pressure on the surface layer is independent of the four-velocity.This is because of the overlap of two suitably defined spaces $V_{+}$ and $V_{-}$  on the patch around P.The hypersurfaces of $V_{+}$ and $V_{-}$ do not intersect,so there are no inner boundaries.We recall that  the norm of four velocity is positive on the space -like hypersurface and negative on the time-like hypersurface.Thus our negative tangential pressure is dependent only on the metric of the hypersurface and the way the normal to the hypersurface bends in the embedding space-time.In other words, how the hypersurface bends in the space in which it is embedded.
\section{Normal Curvature in Curved Space-Time}
\label{sec:2}
To obtain a general relativistic analogue to Kelvin's relation(1),we define the normal curvature  of the Time - like hypersurface $\Sigma$.This $\Sigma$ is contained in a space -time $V_{4}$.$V_{4}$ is the space-time formed by two locally connected regions $V_{+}$ and $V_{-}$ (see Sec.~\ref{sec:1}). We use the Extrinsic curvature of the hypersurface\cite{Hawking1995}.This is because the extrinsic curvature by its definition measures the bending of the hypersurface $\Sigma$ in the space-time in which it is embedded.For static spherically symmetric configurations and for weak fields a delta - like negative tangential pressure coincides with the classical surface tension- \cite{Schmidt1984} and \cite{Taub1980}.This $V_{+}$ space is a hyperboloid sheet curved upwards on the surface.Also the points in its neighborhood lie on the outermost region of the source.Using the masslessness of the surface layer,we ensure that the tangential pressure on it is negative.We define the surface tension on the outermost region of the source in terms of the  negative tangential pressure on the massless hypersurface  of the $V_{+}$ space. On this hypersurface the energy momentum tensor is $T^{ki}=\tau^{ki}.\delta_{\Sigma}$ \cite{Taub1980}.
\begin{equation}
\tau_{\alpha \beta}= g^{\mu \nu}K_{\mu \nu}S_{\alpha \beta}
\end{equation}
Here $\tau_{\alpha \beta}$ is the negative tangential pressure on the hypersurface of  $V_{+}$.$V_{+}$ space has  time-like  hypersurface ,the neighbourhood points of which lie on the outermost region of the source. Here $g_{\mu \nu}$ is the metric on this hypersurface.Also $a_{ij}^{+}$ is the metric coefficient of $V_{+}$space and  $a_{ij}^{-}$ is the metric coefficient of $V_{-}$ space.  Here $S_{\alpha\beta}>0 $ \\ 
Therefore H the analogue of classical mean curvature in Kelvin's relation (eq.1) here reads:

\begin{equation}
 H 
= g^{\mu\nu}K_{\mu\nu}
\end{equation}\\
In our case, H is the normal curvature of Hypersurface of $V_{+} (x^{\mu})$  where:$ x^{\mu}\;(s)\;\mu\;=\;0,2,3$  (see. Appendix I).
$K_{\mu\nu}$ is the extrinsic curvature of the hypersurface.$K_{\mu\nu}$ is also called the \textbf{Second Fundamental form} of the sub- manifold i.e the hypersurface \cite{Caroll2004}.We know from Differential geometry that the inner product of \textbf{Second fundamental form} and \textbf{metric} will give the Normal curvature of the sub-manifold which is the hypersurface here.The normal curvature, therefore like the \textbf{Second fundamental form} and the \textbf{metric}  just depends on the point where we find the normal curvature and the direction of the tangent drawn to the curve at that point.It is a scalar invariant and its sign is positive or negative. This depends on whether the displacement vector between  the point of contact of the two enveloping surfaces with the hypersurface and a neighbourhood point lies on one or the other side of the tangent plane drawn at the point of contact.\\
Recall that the enveloping manifold $V_{4}$ is formed of two locally connected regions $V^{+}$ and $V^{-}$.It  is a Hyperboloid with a thin  overlapping region which represents the hypersurface of the $V_{+}$ manifold(see fig.~\ref{hyperbole}).The normal curvature is taken at any point P on the hypersurface. This is the point of contact of the  hyperboloid sheets  with the hypersurface of $V_{+}$ space.The $V^{+}$ space has  positive normal curvature.This is because it is a hyperboloid sheet curved upwards. The $V^{-}$ space has negative  normal curvature.This is because it is a hyperboloid sheet curved downwards.
The extrinsic curvature  represents how the hypersurface is embedded in the enveloping surface  $V_{4}$.It is given as:
\begin{equation} K_{\mu\nu} = \frac{1}{2}L_{n}P_{\mu\nu}\end{equation}\\
where $L_{n}$ is  the Lie -derivative along the normal vector field to the hypersurface \cite{Caroll2004}.The Tensor $P_{\mu\nu}$ is the projection tensor defined on the hypersurface $\Sigma$ of $V_{+}$ space.It projects any vector in the enveloping manifold $V_{4}$ tangent to the hypersurface.\\
\begin{equation}
P_{\mu\nu} = g_{\mu\nu}  - \sigma n_{\mu}n_{\nu} \end{equation}\
Here $n_{\mu}$ is the unit normal vector to the hypersurface such that $n_{\mu}n^{\mu} =\sigma $.If we choose $n_{\mu}$ ,the normal vector field to the hypersurface in Gaussian  co-ordinates such that it is geodesic everywhere in the neighbourhood of the hypersurface.Then we can write extrinsic curvature $K_{\mu\nu}$ as:
\begin{equation}K_{\mu\nu} =\frac{1}{2}L_{n}g_{\mu\nu} \end{equation}\

Let us consider a hypersurface $\Sigma$ with co-ordinates $x^{\mu}$ ; $\mu = 0,2,3$ which is orthogonally cut by co-ordinate lines of parameter $x^{1}$.The parameter $x^{1}$ measures the arc length along the geodesics from $x^{1} =0$ in the orthogonal direction to it.The parameter $x^{1}$ exists in the immediate neighbourhood of the hypersurface $x^{1}=0$.We assume that these geodesics don't cross each other in the immediate neighbourhood of the hypersurface. \
Since the hypersurface coordinates have no dependence on  orthogonal vector field $\frac{\partial}{\partial x^{1}}$ ,therefore the vector field $\frac{\partial}{\partial x^{1}}$ generates a one-parameter family of symmetries of the metric $g_{\mu\nu}$.(We use  $g_{\mu\nu}$ instead  of $g_{\mu\nu}^{+}$ to represent the metric coefficient of the hypersurface of $V_{+}$ space.Therefore we can write:\
\begin{equation}L_{n}g_{\mu\nu}=0\end{equation}\
\begin{equation}\nabla_{\mu} n_{\nu} + \nabla_{\nu}n_{\mu} =0\end{equation}\
Here eq 10 is the Lie - derivative of the metric of the hypersurface with respect to the normal vector field to it ,which is zero.Eq.11  is the co-variant derivative of the normal vector field.Since the orthogonal trajectories to the hypersurface are taken to be geodesics everywhere,
Therefore we can write:\
\begin{equation}\frac{\partial}{\partial x^{1}}g_{\mu\nu}=0\end{equation}
The  (eq 7) above  is a more general case.This is when the normal vector field is not geodesics everywhere in the neighbourhood of the hypersurface.In our case,we use the Gaussian normal coordinates(see Sec.~\ref{sec:4}) with the normal vector field  geodesic everywhere in the neighbourhood of the hypersurface (eq 7) can be written as below:\
\begin{equation}K_{\mu\nu} = \frac{1}{2}L_{n}g_{\mu\nu}\end{equation}\
  When the integral curves of the normal vector field $n^{\mu}$ are geodesics ,the lie-derivative of the second term in the Projection operator i.e $n^{\nu}\nabla_{\nu}n^
{\mu}$ which is the acceleration goes to zero.\
So,from (eq 11) we can write:\
\begin{equation}K_{\mu\nu}=\frac{1}{2}L_{n}g_{\mu\nu}= \nabla_{\mu}n_{\nu}\end{equation}\
which for the special case of gaussian normal coordinates where the orthogonal vector field is the killing vetor field for the metric of the hypersurface  reduces to zero.But its value varies from reference to reference.\
Most generally the definition of extrinsic curvature i.e.$k_{\mu\nu}$ is:
\begin{equation}K_{\mu\nu}=\nabla_{\mu}n_{\nu} - \sigma n_{\mu}n_{\nu}\end{equation}\
Now going back to our definition of $H$ in equ $6$.In our case it is the normal curvature of a given point P on the hypersurface.From equ's $6$ and $13$ we can write:\
\begin{equation}H = g^{\mu\nu}\nabla_{\mu}n_{\nu}\end{equation}\
Using metric compatibility on the hypersurface we can write:\
\begin{equation}H = \nabla_{\mu}n^{\mu}\end{equation}\
The term$\nabla_{\mu}n^{\mu}$ in the above eq. measures the divergence of unit normal to the surface layer in its $V_{+}$  enveloping space.Therefore ,$\tau_{\alpha\beta}$  which is the negative tangential pressure on this hypersurface can now be written as:\
\begin{equation}\tau_{\alpha\beta}= \nabla_{\mu}n^{\mu}S_{\alpha\beta}\end{equation}\
We can prove in Riemannian geometry that negative of the divergence of unit normal in the enveloping space isthe mean curvature of the hypersurface(see Sec.~\ref{sec:5}).\
Therefore eq.$17$ becomes :
\begin{equation}\tau_{\alpha\beta}= HS_{\alpha\beta}\end{equation}\
\begin{center}$\tau_{\alpha\beta}= -$[mean curvature of hypersurface]$S_{\alpha\beta}$\end{center}
\section{Delta -like Energy flows across the Surface}
\label{sec:3}
\subsection{Surface Tension and Energy-Momentum of the Surface Layer}
\label{subsec:3.1}
The orthogonal vetor field $n^{\mu}$ which generates a one-parameter family of symmetries of the metric $g_{\mu\nu}$ of the surface layer.(see Sec. ~\ref{sec:2}).This orthogonal vector field $n^{\mu}$ is a killing vector for the surface tension tensor $S_{\alpha\beta}$. This is because the surface tension in our case is independent of the four-velocity.It is proportional only to the metric of the surface layer.Therefore the Lie -derivative of the surface tension tensor along the orthogonal vector field should be zero : 
 \begin{equation}L_{n}S_{\alpha\beta}= 0\end{equation}\
 The Lie-derivative of the surface tension tensor $S_{\alpha\beta}$ evaluated in the gaussian normal coordinates (see sec.~\ref{sec:2}) is as follows:
 \begin{equation}L_{n}S_{\alpha\beta}= n^{\sigma}\frac{\partial}{\partial x^{\sigma}}(S_{\alpha\beta}) + \frac{\partial}{\partial x^{\alpha}}(n^{\lambda})S_{\lambda\beta} + \frac{\partial}{\partial x^{\beta}}(n^{\lambda})S_{\alpha\lambda}\end{equation}\
 Recall $x^{1}$ is the parameter that varies along the integral curves of the vector field $n^{\mu}$ (see sec.~\ref{sec:2}).The vector field $n^{\mu}$ is  orthogonal to the surface layer.Note that the surface layer here is the hypersurface of $V_{+}$ enveloping space. then in this particular choice of coordinates,the vector field $n^{\mu} = (\frac{\partial}{\partial x^{1}})^{\mu}$ has components $n^{\mu} = (0,1,0,0)$.Here$x^{\mu} ; \mu = 0,2,3 $ are the coordinates on the surface layer i.e. hypersurface $\Sigma.$ of $V_{+}$ space.
 Therefore ,we write:
 \begin{equation}L_{n}S_{\alpha\beta}= \frac{\partial}{\partial x^{1}}(S_{\alpha\beta}) + \frac{\partial}{\partial x^{\alpha}}(1)S_{\lambda\beta} + \frac{\partial}{\partial x^{\beta}}(1)S_{\alpha\lambda}\end{equation}
 The second and the third terms vanish and we write:
 \begin{equation}\frac{\partial}{\partial x^{1}}(S_{\alpha\beta})= 0\end{equation}
 In (eq 1) ,we give example to prove that the surface tension has the dimensions of intrinsic pressure.We know that the integral I of a scalar function$\phi$ over a n-manifold is:
 \begin{equation}I= \int \phi\sqrt{g} d^{n}x   \end{equation}
 Here $g$ is the determinant of the metric on the manifold which is n-fold here.We write the integral of the scalar function $S_{\alpha\beta}dx^{\alpha}dx^{\beta}$ along any one dimension of the hypersurface below.The energy $E_{\Sigma}$ of the hypersurface,which is a $3$-manifold here is
 \begin{equation}E_{\Sigma}=\int S_{\alpha\beta}dx^{\alpha}dx^{\beta}\sqrt{g}dx\end{equation}
 or using the metric to raise the indices we write:
 \begin{equation}E_{\Sigma}= \int S_{\alpha\beta}g^{\alpha\beta}dx_{\beta}dx^{\beta}\sqrt{g}dx \end{equation} We interpret the above expression as the energy stored in the hypersurface due to the stretching force of the surface tension along any one dimension of the hypersurface.We can do so here because our negative tangential pressure on the surface has no dependence on the four velocity of the surface layer.This is due to choice of the surface composed of overlap of spaces $V_{+}$ and $V_{-}$.Recall $V_{+}$ has a time-like hypersurface $V_{-}$ has space-like hypersurface. Therefore:\begin{equation} g_{\mu\nu}^{\pm}U^{\mu}U^{\nu}=\mp 1\end{equation}The above eq. implies that the norm of the four-velocity is -1 on the time-like hypersurface of $V_{+}$ space and +1  on the space-like hypersurface of $V_{-}$. This means that the points on the surface layer are stationary.That is the surface tension arises only due to the interactions of such stationary points on the surface layer.We know that the internal interactions between points on a surface are always central.Therfore we can say that the work done by the stretching force of surface tension is stored as energy on this hypersurface.We express the energy of the hypersurface in terms of the energy-momentum tensor of the enveloping space $V_{4}$,which is made up of two locally connected regions $V_{+}$ and $V_{-}$ :
 \begin{equation}\int n_{i}T^{ij}n_{j}\sqrt{g}d^{3}x\end{equation}
 The term $n_{i}T^{ij}$ represents the conserved current on the time - like  hypersurface of $V_{+}$.This is  because $\nabla_{j}n_{i}T^{ij}$ is zero.We can recall that the energy-momentum tensor concentrated on a time-like hypersurface is called the surface layer in General relativity.That is the hypersurface of $V_{+}$ space ,the neighbourhood points of which represent the outermost region of the source is the surface layer here.We use the stokes' theorm and relate the divergence of the vector field to its value on the boundary :
 \begin{equation}\int_{M}d^{n}x\sqrt{g}\nabla_{\mu}V^{\mu}=\int_{\partial M}d^{n-1}y\sqrt{\gamma}n_{\mu}V^{\mu}\end{equation}
 We can say that the expression $\int n_{i}T^{ij}n_{j}\sqrt{g}d^{3}x$ is constant on the boundary.The boundary here is the hypersurface embedded in the $V_{+}$ manifold.
 We use the fact that the energy due to the stretching force of the surface tension is stored on the hypersurface and write:
 \begin{equation}\int S_{\alpha\beta}g^{\alpha\beta}dx_{\beta}dx^{\beta}\sqrt{g}dx=\int n_{i}T^{ij}n_{j}\sqrt{g}d^{3}x\end{equation}
 \begin{center}or\end{center}
 \begin{equation}\int S_{\alpha\beta}g^{\alpha\beta}\sqrt{g}d^{3}x=\int n_{i}T^{ij}n_{j}\sqrt{g}d^{3}x\end{equation}\
 Here $g$ is the determinant of the metric on the hypersurface.From $31$ we write:
 \begin{equation}S_{\alpha\beta}=n_{i}n_{j}T^{ij}g_{\alpha\beta}\end{equation}
  The coordinates on the hypersurface are $x^{\mu}=0,2,3 $ so we write:
 \begin{equation}S_{00} + S_{22} + S_{33} = n_{i}n_{j}T^{ij}(g_{00}+ g_{22} + g_{33}) \end{equation}
 \begin{center}or\end{center}\begin{equation}S_{00} + S_{22} + S_{33} = n_{i}n_{j}T^{ij}\frac{1}{2}(a_{11}^{+} + g_{00}+ g_{22} + g_{33} + a_{11}^{-} + g_{00} + g_{22} + g_{33})\end{equation}
 The metric on the enveloping space $V_{+}$ is $a_{11}^{+}dx_{1}^{2} + g_{\mu\nu}dx^{\mu}dx^{\nu}$(see.Appendix. I). We assumed that the surface $V_{+}$ is a hyperboloid sheet curved upwards and with positive normal curvature  at the point of contact P of the hyperboloid sheet with the hypersurface (see.Sec.~\ref{sec:2}).Therefore $a_{11}^{+} =1$. The metric on the $V_{-}$ embedding surface is $a_{11}^{-}dx_{1}^{2} + g_{\mu\nu}^{-}dx^{\mu}dx^{\nu}$.Recall that $V_{-}$ is the hyperboloid sheet curved downwards.Also $V_{-}$ has negative normal curvature (see.Sec.~\ref{sec:2}).Therefore $a_{11}^{-}=-1$.Thus we  write:
 \begin{equation}S_{\alpha\beta} = n_{i}n_{j}T^{ij}\frac{1}{2}(a_{ij}^{+} + a_{ij}^{-})\end{equation}
 Here $a_{ij}^{+}$ and $a_{ij}^{- }$ are the metric coefficients for the $V_{+}$ and $V_{-}$ spaces respectively.
 We know from  eq.$18$ that the Pressure across the surface will be given by:
 \begin{equation}\tau_{11}= \nabla_{\mu}n^{\mu}S_{11}\end{equation} 
  From eq.$35$ we get:
 \begin{equation} S_{11}=n_{i}n_{j}T^{ij}\frac{1}{2}(a_{11}^{+} + a_{11}^{-}) =n_{i}n_{j}T^{ij}\frac{1}{2}(1 + -1)=0\end{equation}
 Using the result of eq. 23  and eq.12 ,we write the energy flow across the hypersurface as :
 
 \begin{equation}\frac{\partial}{\partial x^{1}}(n_{i}n_{j}T^{ij})= \frac{\partial}{\partial x^{1}}(S_{\alpha\beta}g^{\alpha\beta})=0\end{equation}
 Note that the orthogonal vector field $\frac{\partial}{\partial x^{1}}$ is a killing vector of the metric of the hypersurface of $V_{+}$.
 Now we show that for weak -fields and spherically symmetric configuration with constant radii R,the expression of eq.18 i.e. $\tau_{\alpha\beta}=\nabla_{\mu}n^{\mu}S_{\alpha\beta}$  gives the same result for pressure difference across the surface as eq.1.
 \begin{equation}\delta\tau_{\alpha\beta}= \delta\nabla_{\mu}n^{\mu}S_{\alpha\beta} + \nabla_{\mu}n^{\mu}\delta S_{\alpha\beta}\end{equation}
 The term $\delta S_{\alpha\beta}$ is zero due to eq.$14$.There is no change in surface tension across the surface.Therfore we write :
 \begin{equation}\delta\tau_{\alpha\beta}=\delta\nabla_{\mu}n^{\mu}S_{\alpha\beta}\end{equation}
 For weak fields replace $\nabla_{\mu}n^{\mu}$ with $\partial_{\mu}n^{\mu}$ and write:
 \begin{equation}\delta\partial_{\mu}n^{\mu}= \delta \frac{\partial}{\partial x^{\mu}}
 (n^{\mu})\end{equation}
  $\partial n^{\mu}$ is tangential to the surface.Therefore we  write the above as:\begin{equation}\delta\partial_{\mu}n^{\mu}=\delta\partial_{\mu}(e_{t})= \partial_{\mu}(e_{t}) - \partial_{\mu}(-e_{t})\end{equation}.Here $e_{t}$ is the unit tangent vector to the $V_{+}$ surface,which is a hyperboloid sheet curved upwards and with positive normal curvature at the point of contact P with the hypersurface.In the same way, $-e_{t}$ is the unit tangent vector to the $V_{-}$ surface,which is a hyperboloid sheet curved downwards and with negative normal curvature at the point of contact P with the hypersurface.For a spherically symmetric surface of constant radii R:\begin{equation} \frac{\partial e_{t}}{\partial l}=\frac{\pm 1}{R}\end{equation} Here $\partial l$ is the arc length along the surface.The positive sign is for a surface with positive normal curvature and the negative sign for a surface with negative normal curvature at a common point of contact.We  write
  \begin{equation}\delta\partial_{\mu}n^{\mu}= \frac{\partial l}{\partial x^{\mu}}\delta(\frac{\partial}{\partial l} e_{t})= \frac{\partial}{\partial l}(e_{t}) - \frac{\partial}{\partial l}(-e_{t})\end{equation}
  \begin{equation}\delta\partial_{\mu}n^{\mu}= \frac{\partial}{\partial l}(\frac{l}{R})-\frac{\partial}{\partial l}(\frac{-l}{R}) = \frac{2}{R}\end{equation}
 The term $\frac{\partial l}{\partial x^{\mu}}$ goes to unity for $\partial x^{\mu}\rightarrow 0$.This is because $\partial x^{\mu}$ is tangential to the surface.Therefore  we write eq.18 for weak fields and static spherically symmetric configuration of constant radii as :
 \begin{equation}\delta\tau_{\alpha\beta} = \frac{2}{R}S_{\alpha\beta}\end{equation}
The mean curvature of a hypersurface is the negative of the divergence of unit normal in the enveloping space. ,therefore we express the pressure difference $\delta\tau_{\alpha\beta}$ in terms of the mean curvature for weak fields as:
 \begin{equation}\delta\tau_{\alpha\beta}= \frac{-2}{R}S_{\alpha\beta}\end{equation}
 This reduces to the  expression of eq.1 which is the kelvin's relation between surface tension S,pressure difference $\Delta P$ and mean curvature H.
 \subsection{Action : Non-compact Geometries with no inner boundaries} \label{subsec:3.2}
\begin{figure}[h!]
 \centering{
\includegraphics[width=3in]{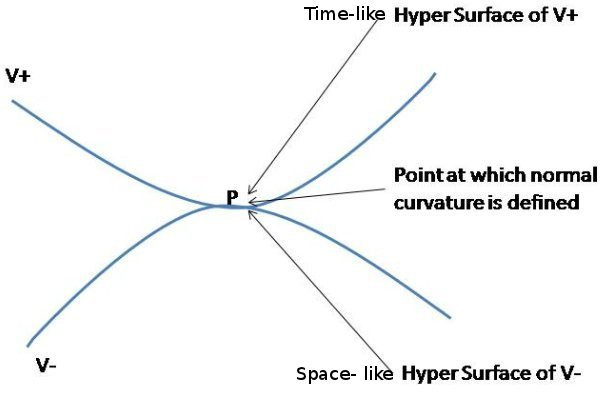}}
\caption{$V_{4}$ space formed by the overlap of $V_{+}$ and $V_{-}$ spaces}
\label{hyperbole}
\end{figure}

 We start with the covariant Lorentzian action \cite{Hawking1995}for a metric g and generic matter fields $\phi$:\begin{equation} I(g,\phi)=\int_{M}[\frac{R}{16\Pi}+L_{m}(g,\phi)] + \frac{1}{8\pi}\oint_{\partial M}K\end{equation}where R is the scalar curvature of g,$L_{m}$ is the matter lagrangian,and K is the trace of the extrinsic curvature of the boundary.The surface term is required so that the action yields the correct equations of motion subject only to the condition that the three metric and matter fields on the boundary remain fixed.The action (eq.48)is well defined for spatially compact geometries,but diverges for non-compact ones.To define the action for non-compact geometries,one must choose a reference background $g_{o}\phi_{o}$.We require that this background be a static solution to Einstein's equations.The physical action is then the difference\begin{equation}I_{p}(g,\phi)\equiv I_{p}(g,\phi) -I_{p}(g_{o},\phi_{o})\end{equation}$I_{p}$ is finite for a class of fields g,$\phi$
which asymptotically approach $g_{o},\phi_{o}$.For Asymptotically flat space-times,the appropriate background is flat space with zero matter fields,and the action of (eq.48) reduces to the familiar form of the gravitational action\begin{equation}I_{p}(g,\phi)=\int_{M}[\frac{R}{16\pi}+L_{m}] + \frac{1}{8\pi}\oint_{\partial M}(K - K_{o})\end{equation}where $K_{o}$ is the trace of the extrinsic curvature of the boundary embedded in flat space time.However,when matter (or a cosmological constant)is included,one may wish to consider spacetimes which are not asymtotically flat.In this case one cannot use flat space as the background,and one must use the more general form of the action (eq.49).For the geometry of the kind discussed in (see Fig.~\ref{hyperbole}), we propose an action \begin{equation}I=\int_{M}[\frac{R}{16\pi}+ L_{m}] +\oint_{\partial M}\frac{1}{8\pi}\nabla_{\mu}n^{\mu}S_{\alpha\beta}g^{\alpha\beta}\end{equation}where the term$\nabla_{\mu}n^{\mu}$ in general represents the sum of the normal curvatures in n-mutually orthogonal directions ,$S_{\alpha\beta}g^{\alpha\beta}$ represents the the surface tension.In the action of(eq.51),when the boundary surface will have maximum normal curvature,the contribution from the surface tension to the energy of the hypersurface will be minimal.When the boundary surface has minimal contribution from the normal curvature,the surface tension contributes maximum to the energy of the hypersurface.This happens in the same sense as in a classical bubble having fixed pressure difference,which depends on the mean curvature and the surface tension.
If the contribution of the curvature decreases,it is made up by the surface tension and vice versa.The decrease of surface tension with higher curvature for classical liquid-interface has been studied in \cite{Renardy2002},\cite{Pawar1999}.We define a reference background for calculating the boundary term.For a general case ,this is a hypersurface whose sum of the normal curvatures in all its n-directions is zero,i.e. the hypersurface is minimal.The physical action for the boundary surface is\begin{equation}\oint_{\partial M}\nabla_{\mu}n^{\mu}S_{\alpha\beta}g^{\alpha\beta}=\oint_{\partial M}[\nabla_{\mu}n^{\mu}S_{\alpha\beta}g^{\alpha\beta}] - \oint_{\partial M}[\nabla_{\mu}n^{\mu}S_{\alpha\beta}g^{\alpha\beta}]_{0}\end{equation}Integrating the first term on the R.H.S. of (eq.52)\begin{equation}\oint_{\partial M}[\nabla_{\mu}n^{\mu}S_{\alpha\beta}g^{\alpha\beta}]=\nabla_{\mu}n^{\mu}\oint_{\partial M}S_{\alpha\beta}g^{\alpha\beta}\end{equation} where$\nabla_{\mu}n^{\mu}$ is the sum of normal curvatures ,which is a scalar invariant.Recall from (eq.32),the integral of the surface tension term gives the energy of the hypersurface.Note that the manifold in our case(see Fig.~\ref{hyperbole})and (see Sec.~\ref{sec:1}) is formed by the overlap of two spaces $V_{+}$ and $V_{-}$ respectively.So,there are no inner boundaries.Also recall that the hypersurfaces are assumed to be massless.The tangential pressure on them is independent of the four-velocity by construction.In the second integral,the term $[\nabla_{\mu}n^{\mu}S_{\alpha\beta}g^{\alpha\beta}]_{0}$ is zero.Therefore $[\nabla_{\mu}n^{\mu}]_{0}$ is zero,as the sum of normal curvatures for a minimal surface, taken as reference background vanishes.Therefore the second integral will generate a constant term. This constant term is the boundary contribution of the reference backbround.

 \section{Conclusions:}
 \label{conclusion}
 Our purpose is to represent surface tension in a more generalized and natural geometric way.We express the tangential presssure in terms of normal curvature of the surface layer.Normal curvature is defined using the extrinsic curvature of the surface layer.We consider  a very thin surface layer.Using a time-like hypersurface and a space-like hypersurface overlapping its inner patch, the tangential pressure on the surface is defined to be independent of the four-velocity.This has been reformulated and simplified compared with the result of \cite{Schmidt1984}.Therefore,the surface tension is also independent of four-velocity of the surface layer.It only depends upon the metric of the surface layer and how the hypersurface bends in the space-time in which it is embedded.This way of representation can be claimed to be more naturally geometric.Normal curvature  is defined only in terms of the point where it is calculated on the hypersurface.It does not require any definition of boundary on the hypersurface as in.It is more generalized.This is because the boundary is free from constraints to be imposed on the Four-velocity of the hypersurface as in \cite{Schmidt1984}.The  enveloping space is m-dimensional and its subspace i.e the hypersurface here, is n-dimensional such that$m=n+1$.This implies there is only one unique linearly dependent unit normal which can be defined on the hypersurface.Therefore there is only one unique normal curvature  of  subspace in such cases.Or all the curves through the point P on the hypersurface will have the same normal curvature till the sense of the tangent at point P changes.We assume that the work done by surface tension in stretching of surface is stored as energy on the hypersurface.This is due to internal interaction between stationary points  on a surface layer always being central. We calculate the energy of a very thin hypersurface from the energy-momentum tensor of the enveloping space.We get an expression for surface tension tensor$S_{\alpha\beta}$ in terms of the energy-momentum tensor $T^{ij}$ of the enveloping manifold.The energy-momentum tensor$T^{ij}$ for the enveloping space is evaluated from the Einstein's equations for the source.We also propose a modified action term for the boundary surface.This is based on the contributions both from the extrinsic curvature and the surface tension in the boundary term.This is useful in cases where we cannot chose a flat space-time as a reference background.We propose to choose a reference background ,which is minimal,i.e. the sum of normal curvatures in all its n-mutually orthogonal directions is zero.We will calculate the action in the Hamiltonian formalism for the new form boundary term in the following paper.
 \section{Appendix. I}
 \label{sec:4}

Let the metric on time -like hypersurface be :
\begin{equation}
ds^{2} = -(d x^{0})^{2}+ R^{2}d\Omega^{2}
\end{equation}

where $ x^{0}$ is the proper time  and $ x^{\mu}$ ,$\mu = 0,2,3$ be co-ordinates in $\Sigma$.Here $d\Omega^{2}$ is identical to $d\theta^{2} + \sin^{2} \theta d\phi^{2}$ and $\xi^{i} i=0,1,2,3$ those for the enveloping space $V_{4}$.
The embedding of $\Sigma \subset V_{4}$ is performed by function $\xi^{i}(x^{\alpha})$ whose derivatives are $e^{i}_{\alpha}$ equal to $\frac{d\xi^{i}}{d\xi^{\alpha}}$ equal to $ \xi_{i}^{i},\alpha $.
$\Sigma$ divides $V_{4}$ atleast locally into two connected components $V_{+}$ and $V_{-}$ and the normal $n_{i}$ is defined by:\begin{equation}
n_{i}n^{i}=1\end{equation} for a time-like hypersurface and \begin{equation}
 n_{i}n^{i}=-1\end{equation} for a space-like hypersurface.
 \begin{equation}n_{i}e^{i}_{\alpha} = 0\end{equation}We define normal $n_{i}$ in this way and ensure that the tangential pressure on the hypersurface is negative both for a time -like and a space -like hypersurface.This is because a very thin hypersurface is assumed to be massless. The orthogonal normal to the hypersurface of $V_{+}$ space points outward.The orthogonal normal to the hypersurface $V_{-}$ space points inward.We assume a special co-ordinate system
\begin{equation}
x^{i}\;= \xi^{\mu}
\end{equation}Then $x^{1}\; = 0$ is a co-ordinate hypersurface.The geodesics cutting orthogonally the hypersurface  $x^{1}\;=0$ are co-ordinate lines of parameter $x^{1}$.Then the parameter $x^{1}$ measures the arc length along these geodesics from $x^{1}\;=0$.Let $Q^{i}$ be the tangent vector to a geodesic which is orthogonal to the co-ordinate hypersurface $x^{1}\;=0$ in the $V_{+}$ enveloping space.Then $Q^{i}\;=\;(0,dx^{1},0,0)$ and the length of $Q^{i}$ is given by \begin{equation}
Q^{2}=a_{ij}^{+}Q^{i}Q^{j}=a_{11}^{+}Q^{1}Q^{1} + g_{\mu\nu}Q^{\mu}Q^{\nu}\end{equation}where $g_{\mu\nu}$ is the metric on the hypersurface of $V_{+}$ space..Therefore\begin{equation} Q^{2}=dx^{1}dx_{1}
=a_{11}^{+}dx^{1}dx^{1}
= (dx^{1})^{2}
 =a_{11}^{+}=1\end{equation}
Let $P^{i}$ be the tangent vector to the hypersurface $x^{1}=0$.Here $P^{i}$ equal to                                                        $(dx^{0},0,
dx^{2},dx^{3})$.If $P^{i}$ is orthogonal to $Q^{i}$ then\begin{equation}
a_{ij}^{+}Q^{i}P^{j}= a_{1j}^{+}Q^{1}P^{j}=0\end{equation}
\begin{equation}a_{1j}^{+}P^{j}=0  [Q^{1}\neq 0;P^{j}\neq 0;(j=0,2,3)
\Rightarrow a_{1j}^{+}  =0]\end{equation}The term$a_{1j}^{+}= 0$ implies that the orthogonal vector field tangent to the geodesics ,which is parametrized by$x^{1}$ is orthogonal to the family of hypersurfaces diffeomorphic to the given hypersurface.From the differential equation of geodesics in the enveloping space $V_{+}$,
\begin{equation}
 \frac{d^{2}x^{i}}{d(x^{1})^{2}} + \Gamma^{i}_{11}\frac{dx^{1}}{ds}\frac{dx^{1}}{ds} =0\end{equation}
 \begin{equation} 
 \Gamma^{i}_{11}=\frac{1}{2}(a^{ij,+})[11,j]=0  
 therefore [\frac{d^{2}x^{i}}{(dx^{1})^{2}} =0  ;i=0,1,..,3] 
 \end{equation}
 Therefore everywhere on the co-ordinate hypersurface we write:
 \begin{equation}
a_{11}^{+}=1; a_{1j}^{+}=0 ;j\neq 1[\frac{\partial a_{1j}^{+}}{\partial [x^{1}]}=0 ; a_{1j}^{+}=0;
\frac{\partial a_{1j}^{+}}{\partial x^{1}}=0 ;j \neq 1]\end{equation}
The line element in the enveloping space $V_{+}$ is : \begin{equation}
ds^{2}=(dx^{1})^{2}\;+ g_{\mu\nu}dx^{\mu}dx^{\nu} [\mu =\nu \neq 1] \end{equation}
By the same analysis the line element in the enveloping space $V_{-}$ :
\begin{equation}
ds^{2}= -(dx^{1})^{2}+ g_{\mu\nu}^{-}dx^{\mu}dx^{\nu} [\mu =\nu \neq 1]
\end{equation}Thus we see that the manifolds of the two enveloping spaces are different. The hypersurface is time-like in $V_{+}$ space and space-like in $V_{-}$ space.The definition of normal curvature is only dependent on the point of contact of this surfaces and the sense of the normal.Therefore the world lines of the point of contact P which is a stationary point in the two enveloping spaces $V_{+}$ and $V_{-}$ will point in mutually opposite directions in the frame of the static spherically symmetric source.Therefore their normal curvatures have the opposite signs.The massless sharacter of the hypersurfaces in any case ensures that the tangential pressure is negative, both for the time-like  and the space -like hypersurface. 
\section{Appendix.II}
\label{sec:5}
Let $n^{i}$ in $V_{n+1}$ having $x^{i}$ such that $i=1,2,..,n+1$, be the unit normal to the hypersurface  $V_{n}$  having coordinates $\xi^{\mu}: \mu =1,2...,n$.Let $L^{i}_{(h)}(h=1,2..,n)$ in $V_{n+1}$ be the  n unit tangents to n congruences $L_{(h)}$ in the orthogonal ennuple of the hypersurface. $V_{n}$.Since $L^{i}_{(h)}$ is orthogonal to $n^{i}$,we write:\begin{equation}L^{i}_{(h)}n_{i}=0 [i=1,2...,n+1;h=1,2...,n]\end{equation}Differentiating covariantly with respect to$x^{j}$,we get\begin{equation}L^{i}_{(h)}n_{i;j} + L^{i}_{(h);j}n_{i}=0\end{equation}Multiplying by $L^{j}_{(h)}$ gives:\begin{equation}L^{i}_{(h)}n_{i;j}L^{j}_{(h)} + L^{i}_{(h);j}n_{i}L^{j}_{(h)} =0\end{equation}\begin{equation}\Rightarrow L^{i}_{(h);j}L^{j}_{(h)}n_{i} =-n_{i;j}L^{i}_{(h)}L^{j}_{(h)}\end{equation}\begin{equation}-n_{i;j}L^{i}_{(h)}g^{ij}L_{i(h)}=-n^{j;j}\end{equation}The R.H.S. of eq.57 represents the divergence of the unit normal in the enveloping space $V_{n+1}$.The L.H.S. term can be interpretted as the sum of the normal curvatures for the n mutually orthogonal directions of the hypersurface.we know from Riemannian geometry that if$V{m}$ is a m-dimensional space and $V_{n}$ is a n-dimensional subspace of $V_{m}$ i.e. $m>n$,then there exists $m-n$ linearly independent unit normals to the subspace $V_{n}$.Therefore in this case as the subspace i.e. hyperspace has one dimension less than the enveloping space ,so there exists only one linearly independent normal to the hypersurface.In Riemannian geometry the sum of the normal curvatures in n-mutually orthogonal directions of the hypersurface is the mean curavture of the hypersurface.Therefore eq.57 shows that the mean curvature of a hypersurface is negative of the divergence of unit normal in the enveloping space.

\begin{acknowledgements}
 The suggestions of Prof.Alexei.Toporensky. during the preparation of manuscript are greatly acknowledged.The guidance of Dr.Mir. Faizal in scientific preparation of the manuscript are also acknowledged.We wish to thank Md.Mahfoozul Haque for valuable advice in formatting of the manuscript in Latex. Also we are thankful to Zaheer and Debendu for reading the manuscript and other valuable discussions.
\end{acknowledgements}

\end{document}